\documentclass[runningheads,a4paper]{llncs}

\usepackage{graphicx}
\usepackage{fancyvrb}
\usepackage{setspace}
\usepackage{geometry}
\usepackage{float}
\usepackage{setspace,supertabular}
\usepackage{longtable}
\usepackage[usenames]{color}
\usepackage{verbatim}
\usepackage{subfigure}
\usepackage{amsmath}
\usepackage[table]{xcolor}
\usepackage{lscape}
\usepackage{changepage}
\usepackage{tabularx}

\graphicspath {{./} {Figures/}}

\newcolumntype{H}{>{\columncolor{black}\color{white}}c}

\graphicspath {{./} {Figures/}}

\newcommand{\keywords}[1]{\par\addvspace\baselineskip
\noindent\keywordname\enspace\ignorespaces#1}

\begin{document}
\title{Artifact Centric Business Process Management\\Logging Schema}

\titlerunning{Artifact Centric Business Process Management Logging Schema}

\author{Mani Baradaran-Hosseini\inst{1}}
\institute{University of New South Wales, Sydney, Australia\\
\email{manibh@cse.unsw.edu.au}}

\authorrunning{Mani Baradaran-Hosseini}

\maketitle

\begin{abstract}

Understanding the evolution of business artifacts will enable business analyst to discover more insight from process execution data. In this context, describing how the artifacts are wired, helps in understanding, predicting and optimizing the behavior of dynamic processes. In many cases, however, process artifacts evolve over time, as they pass through the business's operations. Consequently, understanding the evolution of artifacts becomes challenging and requires analyzing the provenance of business artifacts.
In this paper our aim is to analyze and classify existing challenges in artifact centric business processes. We propose to extend \emph{Provenance} techniques to artifact centric BPMs in order to perform cross cutting concerns on BPMs. Provenance is pre-requirement of addressing cross cutting concerns, which will provide information regarding artifact instance creation and its evolution during its life cycle. Due to dynamic nature of dynamic processes and declarative structure of Artifact Centric BPM systems, it's vital to make sure how an artifact instance actually executed and evolved during its processing in run time.

\end{abstract}

\newpage

\section{Introduction}

Dynamic processes have flexible underlying process definition where the control flow between activities cannot be modeled in advance but simply occurs during run time~\cite{DBLP:conf/bpm/BeheshtiBNS11}. Understanding the evolution of business artifacts will enable business analyst to discover more insight from process execution data. In this context, describing how the artifacts are wired, helps in understanding, predicting and optimizing the behavior of dynamic processes. In many cases, however, process artifacts evolve over time, as they pass through the business's operations~\cite{beheshti2012organizing}. Consequently, understanding the evolution of artifacts becomes challenging and requires analyzing the provenance~\cite{DBLP:journals/corr/abs-1211-5009} of business artifacts.
\\\\
Many artifact centric approaches used business artifacts, that combine data and process in a holistic manner, as the basic building block. Some of these works used a variant of finite state machines to specify lifecycles. Some theoretical works explored declarative approaches to specifying the artifact lifecycles following an event oriented style.
\\\\
Some other works focused on modeling/querying `artifact centric' processes, where a document-driven framework used to model BPM systems through monitoring document lifecycle. A self-learning mechanism can be used for determining the type of the document in business processes through combining process information and document alignment. In these approaches, the document structure is basically predefined.
\\\\
Artifact-centric workflows~\cite{hull2009facilitating} use a predefined process model which describe the lifecycle of the documents. Some other works~\cite{DBLP:conf/caise/BeheshtiBN13}, focused on modeling and querying techniques for business artifacts, received high interest in the research community.
In such models, actors, activities, artifacts, and artifact versions are first class citizens, and the evolution of the activities on artifacts over time is the main focus.
These models supports timed queries and enables weaving cross-cutting aspects, e.g., versioning and provenance, around business artifacts to imbues the artifacts with additional semantics that must be observed in constraint and querying ad-hoc processes.
\\\\
In this paper our aim is to analyze and classify existing challenges in artifact centric business processes. We propose to extend \emph{Provenance} techniques to artifact centric BPMs in order to perform cross cutting concerns on BPMs. Provenance is pre-requirement of addressing cross cutting concerns, which will provide information regarding artifact instance creation and its evolution during its life cycle. Due to dynamic nature of ad hoc processes and declarative structure of Artifact Centric BPM systems, it's vital to make sure how an artifact instance actually executed and evolved during its processing in run time. Main challenge is, some times it's not clear the way artifact is processed because its ad-hoc. Hence, as there are no fixed process model for Artifact Centric BPMs, like structured or semi-structured process models; it's always important to make sure artifact instance processed in expected way afterwards.
\\\\
The remainder of this paper is organized as follows:
We provide an introduction to business processes in Section~\ref{sec1}.
Section~\ref{sec2} presents the challenges for artifact centric business processes.
In Section~\ref{sec3} we analyze the GSM Life Cycle Meta-model.
In Section~\ref{sec4} we present a motivating example.
In Section~\ref{sec5} we analyze and classify the existing artifact centric research challenges.
Finally, we conclude the paper with a prospect on future work in Section~\ref{sec6}.

\section{Introduction to Business Processes}
\label{sec1}

Business Process Management (BPM) systems are Service Oriented Architecture base software applications; that are introduced to responds to complicated automation requirements of business organizations. In order to implement BPM system for an organization, business analyst has to analyze organizations activities~\cite{DBLP:journals/fuin/AalstV14,DBLP:conf/wise/BeheshtiBNA12,DBLP:conf/apbpm/Aalst13} and model them in the computer system.

In conventional BPM systems, business analysts are modeling company's business logic into chain of processes and tasks including their contextual information in computer systems with respect to organization business goals and models~\cite{DBLP:conf/bpm/BeheshtiBNS11}. In order to model real life business logic, business processes should be identified and WBS\footnote{Work Breakdown Structure} guidelines should be followed to break every business process into smaller atomic sub processes till it can be implemented as an activity or task in BPMS. BPMS usually use technologies like WFMS\footnote{Work Flow Management System} , which provides ability to process chains of petri net like processes and tasks with their transitions in BPM.

The out come of above process modeling and design will be a process model, which not understandable for business stakeholders with no knowledge of BPM Modeling. Hence, it can introduce some confusion and miss understandings, which need to be verified by business stakeholders and managers. From the other hand, recent increasing interest for frameworks for specifying and deploying business operations and processes that combine both data and process as first-class citizens, triggered series of researches on alternative approaches~\cite{hull2011formal,DBLP:journals/topnoc/AalstSW13}.

Thus, a new approach has been developed at IBM, called Artifact based BPM which is based on early idea of Adaptive Business Objects. The main goal is to treat both data and process as first-class citizen and instead of focusing on business processes they will focus on business artifacts~\cite{hull2011business}. Business artifact is referred to business entities that play main rule in organizations activity, and they can evolve over the time (eg: their status and attribute values may change during processing)~\cite{DBLP:conf/caise/BeheshtiBN13}.

\section{Artifact Centric BPMs}
\label{sec2}

Artifact centric approach is focusing on augmented data records, called "Business Artifacts" (simply Artifacts), which corresponds to key main business organization object, and their life cycle and their invocation~\cite{hull2008artifact,DBLP:conf/caise/BeheshtiBN13}. Artifact based BPMs have different characteristics in comparison to conventional BPMs (see Table~\ref{tbl1}).

\begin{table}
\begin{adjustwidth}{0cm}{}
 \caption{Comparison of Artifact based BPMs and conventional BPMs.}
 \centering
  \begin{tabular}{cc}
   \includegraphics[scale=0.82]{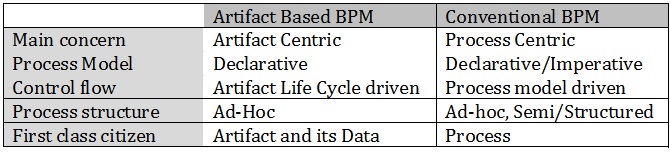}\\
  \end{tabular}
 \label{tbl1}
\end{adjustwidth}
\end{table}

Artifact centric BPMs are inherently declarative process models, process model is not explicitly modeled, instead lifecycle controls the evolution of Artifact during its processing. Thus, as Artifacts can get evolved over a time, Artifact Life Cycle model plays important rule in design of such systems.

As described above Artifacts has life cycle, which shows state of Artifact at each particular point of time. Artifact's life cycle should be finite and has start and ending points, for successful ending and ending with failure. Hence, the life cycle of artifact can be defined by FSM\footnote{Finite State Machine} which can provides the ability of state adaptive access control based on business role an state of artifact~\cite{nandi2005adaptive,DBLP:conf/caise/BeheshtiBN13}. Some artifacts will be short lived (e.g. purchase order) while others can be long lived (eg loggers or monitoring related artifacts)~\cite{hull2008artifact}.

\begin{figure} [t]
\centering
  \includegraphics[scale=1.2]{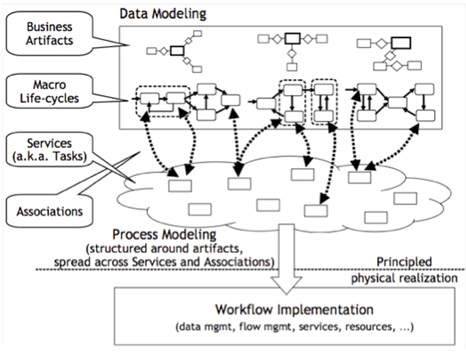}\\
  \caption{Four ``dimensions'' in artifact-centric business process modeling.}\label{fig1}
\end{figure}

In addition, an Artifact type contains both an information model that captures all of the business relevant data about entities of that type; and a lifecycle model to specify the possible ways (states) that artifact can evolve by responding to events and invoking services, including human interaction and activity. Hence, each artifact instance will contain all necessary data and its life cycle meta data during its execution.

Interactions between artifact instances and flow of activity, are supported both through testing of conditions against the artifact instances, and through events coming from an external environment or resulting from changes in artifact instances. The core of the operational semantics is based on the use of rules that are inspired by the Event-Condition-Action (ECA) rules paradigm~\cite{hull2011formal}.

Currently few implementations of Artifact centric BPM engines are existed, Barceluna proto~\cite{hull2011formal} type based on GSM\footnote{Guard Stage Milestone Lifecycle meta model} semantics developed at IBM Watson research center, and EZ-Flow are two implementations of artifact centric BPM engines.

\section{GSM Life Cycle Meta-model}
\label{sec3}

As discussed before, Artifact's Life Cycle is most important part of Artifact Centric BPM design and implementation. Life cycle describes how the state of artifact can change over the time, and governs possible ways that artifact can change and possible transitions from/between each state of artifact~\cite{lohmann2013compliance,DBLP:journals/tsc/Aalst13}. In addition, each Artifact should have at least one final states (means successful processing) and at least one initial state. Hence, using FSM\footnote{Finite State Machine} will make sense to contain all possible states of Artifact and to provide the ability to control the state transitions.

Using FSM can provide the ability of State Adaptive Access Control; which means access level to Artifacts data model via CRUD operations can be controlled with respects to its current state. For example in purchase order scenario, PO can be considered as an Artifact; from the moment that customer places his PO till the PO get approved, customer will have read and update and delete access on PO. When PO get approved customer can not make any further changes, hence will have read only access to his PO only.

Guard-Stage-Milestone (GSM) Life Cycle meta-Model\footnote{Named meta-model to be distinguished from Data Base Model} has been designed and developed at IBM research center based on BEL's approach~\cite{hull2011introducing}. Which, acts as centralized life cycle management engine (system), with focus on governing life cycle of artifacts and their interactions with environment via Event-Condition-Action (ECA) like rule paradigm , with out caring about implementation of services and processes~\cite{hull2011business}. Hence, when system receives external events, the focus is on changes on stages (opened/closed) and milestones (achieved/invalidated) as a result of processing that event.

Handling ECA rules in GSM system are centered on Business Steps or simply B-Step, which corresponds to smallest unit of business-relevant change that can occur within GSM system. Thus, as described before GSM supports the management of business related activities with no concerns about details of execution of activities. Instead, main focus is on changes in life cycle of artifacts as a result of events and b-steps.

In GSM framework, an Artifact Service Center (ASC) is designed to maintain business artifact types and their instances during the runtime. ASC acts like a container of artifact types and instances and in addition, it provides support for SOA interfaces like WSDL/REST to interact with external environment. ASC supports both both-way service call and one-way service call patterns~\cite{hull2011business}.

GSM meta-Model consists of following main components to accomplish its tasks to govern artifact life cycle and its interactions with environment and other artifacts~\cite{hull2011business}:

\begin{itemize}
  \item Information model: integrated view of all artifact related business data during its life cycle. This view provides access to artifact's data model and all of its related data that might be need over life cycle of artifact.
  \item Milestone: Operational objective or target that can be achieved during life cycle of artifact. If milestone implements as Boolean condition, when milestone achieve its target/objective its status/value will become true and if it cannot be achieved status will be false.
  \item Stage: group of activities that can be performed to achieve a milestone owned by the stage. Each milestone represents a way that stage can be completed and at most one milestone of stage can be true at a time. When one of the milestones within stage achieved, stage become inactive (or closed). Because the overall goal of stage execution is to achieve one of its milestones.
  \item Guard: used to controls when stages should become active (or open).
  \item Sentry: used as guards to govern the stages (when they get open or close) and milestones whether they achieved or invalidated.
\end{itemize}

Note: both milestones and guards are controlled in a declarative manner, based on triggering events and/or conditions.

\section{Motivating Example}
\label{sec4}

In this section we use Requisition and Procurement Orders scenario provided at~\cite{hull2011business} to demonstrate how GSM life cycle model will work in real life scenario. In this scenario Manufacturer will receive a Requisition Order (RO) from customer, which may contains multiple Line Items (LI). Next, manufacturer will individually search for its supplier for each Line Item, by sending Procurement Orders (PO) to each supplier. In this example, we will focus on management of orders received from customers RO and to suppliers PO. Thus, its natural to have 3 artifact types in this scenario RO, LI and PO.

\begin{itemize}
  \item RO: each instance manages overall operation of each single received RO.
  \item LI: each instance manages a single line item from list of LIs.
  \item PO: each instance manages a single PO order to supplier.
\end{itemize}

One of the assumptions is supplier can reject a PO at any time before completion and shipment to manufacturer. Thus, the LI of that order must be researched again, and bundled into new PO to be sent to other suppliers. We can define these rules by using combination of milestones and sentries.

\begin{figure} [t]
\centering
  \includegraphics[scale=1.2]{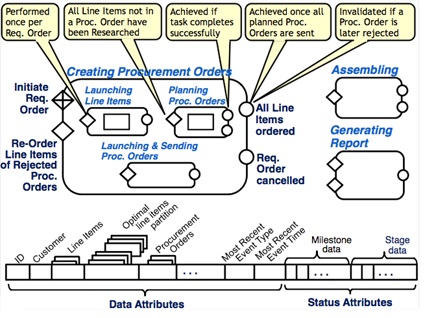}\\
  \caption{Proposed artifact type for RO.}\label{fig2}
\end{figure}

RO artifact is demonstrated in Figure~\ref{fig2}; Milestones are small circles, the diamonds are Guards and rounded corner rectangles are Stages. The diamonds with a cross are "bootstrapping" Guards, which are used to indicate the conditions under which new artifact instances may be created. Rectangles inside stages are Tasks and as you can see stages can be nested. Sentries are suggested in yellow call out boxes. Data Attribute in bottom of Figure~\ref{fig2} contains all business-related data about RO instance. And Status Attribute holes details of current state of milestones (true/false) and stages (open/close) and time stamp of last update time~\cite{hull2011formal}.

To satisfy above assumption, "All Line Items Ordered" milestone will become true or achieved only when all of its planned PO's have been sent out, and will be invalidated if one/more of PO's been rejected by suppliers at anytime before completion and shipment. In case supplier rejected a PO, system will research to other suppliers and send out POs to them. A GSM B-step corresponds to handling of a single incoming event into a GSM system including all achieving/invalidating milestones and opening/closing stages caused by processing on that event~\cite{hull2011formal}. Figure 3; illustrate possible interactions between artifacts related to 3 B-Step.

\begin{figure} [t]
\centering
  \includegraphics[scale=1.2]{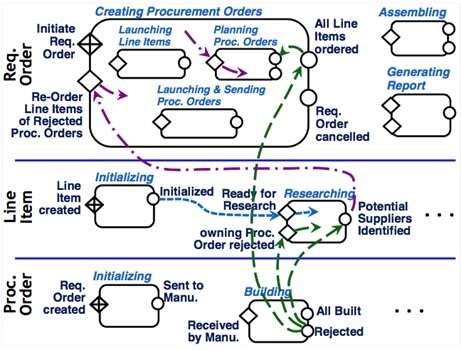}\\
  \caption{Interactions between artifacts.}\label{fig3}
\end{figure}

For example, consider the 5 green arrows from PO milestone called Rejected, this corresponds to scenario that supplier rejected one of the line items. Thus, RO "All Line Items Ordered" will be invalidated even if it's currently true. Thus, now there are some LIs that must be reordered from other suppliers; which means milestone "successful completion" of stage "Planning Proc Orders" will get invalidated as well due to invalidation of "All Line items ordered" milestone. Next, for each LI owned by PO instance which received Reject milestone, "Potential Suppliers Identified" milestone of stage "Researching" will get invalidated (if its true) and "Owning Proc Order rejected" guard will get triggered and its stage (Researching) will be reopened. Next, purple path shows how the control flow will move toward RO artifact to resend the PO requests to new identified suppliers of rejected LI.

\section{Existing Artifact Centric Research Challenges}
\label{sec5}

After briefing on Artifact Centric BPMs and reviewing GSM life cycle model in previous chapter; in this chapter we will briefly focus on current research challenges related to Artifact Centric BPMs.
One of basic challenges in business process modeling is to find a method that business stakeholders like executive managers and analysis, can define business operations in detail. The specification should enhance development of this automation software. At the same time proposed framework should support flexibility, monitoring and reporting. These requirements lead to following research direction and challenges which described in~\cite{hull2008artifact}, please refer to referenced paper for complete list:

\subsection{Model and Views}

Promoting artifact's data as first class citizen will lead to new set of researches to find proper data model and view based on specific behaviors of artifact centric BPMs. traces of such a researches can be find in Adaptive Business Objects~\cite{nandi2005adaptive}, business entities and document driver workflows~\cite{wang2005framework}.  In addition, even thou that data model and storing business process data is open topic for almost a decade (for traditional business processes), there is no one method and model that can be considered as best one after all.

\subsection{Design principles in support of usability and flexibility}

Unlike database systems, which have rich literature and practices in area of database design, in business process world it's still an open challenge to determine best design and approach with respect to their features. Some fundamental questions like: when one has to use procedural approach and when to use declarative approach to specify part of business process? Or how can we measure flexibility among different schemas? Thus, same challenges remains open in artifact centric BPMs.

\subsection{Componentization and Composition}

The Service Oriented Architecture(SOA) design pattern provides capabilities for breaking software into smaller reusable components and services. In this category of researches the focus will be on best practice for designing reusable components for artifact centric BPMs.

\subsection{Monitoring and tuning}

Heart of successful BPM is the ability to monitor the performance of an operation to provide immediate feedback and additional added value details about runtime execution, which can lead to better tuning of system. Artifact centric BPMs provide good support for monitoring systems as artifact instance bundled all required business relevant information with them. In order to perform crosscutting aspects on BPM Systems such as monitoring, auditing or verification; provenance problem has to be addressed first. Thus, proper event logs needs to be generated in order to track/trace state of artifact components.

\section{Provenance and Artifact Centric Approach}

Provenance refers to the documented history of an \emph{immutable} object and often represented as graphs~\cite{DBLP:journals/corr/abs-1211-5817}. The ability to analyze provenance graphs is important as it offers the means to verify data products, to infer their quality, and to decide whether they can be trusted~\cite{DBLP:journals/corr/abs-1211-5009}. In a dynamic world, as data changes, it is important to be able to get a piece of data as it was, and its provenance graph, at a certain point in time. Under this perspective, the provenance queries may provide different results for queries looking at different points in time. Enabling time-aware querying of provenance information is challenging and requires: (i) explicitly representing the time information in the provenance graphs, and (ii) providing timed abstractions and efficient mechanisms for time-aware querying of provenance graphs over an ever increasing volume of data.

In this research our aim is to extend "Provenance" techniques to artifact centric BPMs in order to perform cross cutting concerns on BPMs. As discussed in previous chapter one of current challenges and research topics in Artifact Centric BPMs is monitoring. Provenance is pre-requirement of addressing cross cutting concerns, which will provide information regarding artifact instance creation and its evolution during its life cycle. Due to dynamic nature of ad hoc processes and declarative structure of Artifact Centric BPM systems, it's vital to make sure how an artifact instance actually executed and evolved during its processing in run time. Main challenge is, some times it's not clear the way artifact is processed because its ad-hoc. Hence, as there are no fixed process model for Artifact Centric BPMs, like structured or semi-structured process models; it's always important to make sure artifact instance processed in expected way afterwards.

As discussed in motivation example, Barcelona proto type is one artifact centric BPM system, which is using GSM Life cycle meta-model developed by IBM research center. In this proposal, we aim to propose a logging schema, which later can be used to address provenance problem for this BPM engine.
We will also show that how this feature can be used for enriching crowd computing graphs~\cite{DBLP:conf/apweb/AllahbakhshIBBBF13},e.g., establishing weighted edges between requesters and workers in a crowdsourcing~\cite{DBLP:conf/colcom/AllahbakhshIBBBF12,DBLP:journals/compsec/AllahbakhshIBBFB14} graph.

Required logging schema needs to provide all required information regarding artifact evolution during its runtime in order to address provenance problem. Such a logging schema can utilize the notion of W7~\cite{ram2009new} to capture all required aspects of artifact evolution during its life cycle. For example it may need to facilitate following correlations in Barcelona engine:

\begin{itemize}
  \item Incoming event <-> Artifact (Artifact initialization)
  \item Artifact <-> Artifact (association between artifact)
  \item Guards <-> Stages
  \item Stages <-> Milestones
  \item Guards <-> Milestones
\end{itemize}

\section{Conclusion and Future Work}
\label{sec6}

The continuous demand for the business process improvement and excellence has prompted the need for business process analysis in the enterprise. Recently, business world is getting increasingly dynamic as various technologies such as Internet and email have made dynamic processes more prevalent. Following this, the problem
of understanding dynamic BPs execution has become a priority in the enterprise. In particular, analyzing the evolution of business artifacts is a crucial requirement for many companies in order to understand and improve their business. In this paper we provided an overview of the challenges in artifact centric processes. In our continuous work, our aim is to extend provenance techniques to artifact centric BPMs in order to perform cross cutting concerns on BPMs. Due to dynamic nature of dynamic processes and declarative structure of Artifact Centric BPM systems, it's vital to make sure how an artifact instance actually executed and evolved during its processing in run time. We plan to leverage the semantic techniques proposed in~\cite{DBLP:conf/springsim/BeheshtiM07} to organize the network between artifacts and to propose an economic-based service for provenance discovery through study the existing methods of provenance discovery and using semantic information and Service Oriented Architecture (SOA).

\bibliographystyle{plain}
\bibliography{Biblio}

\begin{thebibliography}{10}

\bibitem{DBLP:conf/apbpm/Aalst13}
Wil~M.P. Aalst.
\newblock Process cubes: Slicing, dicing, rolling up and drilling down event
  data for process mining.
\newblock In Minseok Song, Moe~Thandar Wynn, and Jianxun Liu, editors, {\em
  Asia Pacific Business Process Management - First Asia Pacific Conference,
  {AP-BPM} 2013, Beijing, China, August 29-30, 2013. Selected Papers}, volume
  159 of {\em Lecture Notes in Business Information Processing}, pages 1--22.
  Springer, 2013.

\bibitem{DBLP:journals/tsc/Aalst13}
Wil~M.P. Aalst.
\newblock Service mining: Using process mining to discover, check, and improve
  service behavior.
\newblock {\em {IEEE} T. Services Computing}, 6(4):525--535, 2013.

\bibitem{DBLP:journals/topnoc/AalstSW13}
Wil~M.P. Aalst, Christian Stahl, and Michael Westergaard.
\newblock Strategies for modeling complex processes using colored petri nets.
\newblock {\em T. Petri Nets and Other Models of Concurrency}, 7:6--55, 2013.

\bibitem{DBLP:journals/fuin/AalstV14}
Wil~M.P. Aalst and H.~M.~W. Verbeek.
\newblock Process discovery and conformance checking using passages.
\newblock {\em Fundam. Inform.}, 131(1):103--138, 2014.

\bibitem{DBLP:conf/colcom/AllahbakhshIBBBF12}
Mohammad Allahbakhsh, Aleksandar Ignjatovic, Boualem Benatallah,
  Seyed{-}Mehdi{-}Reza Beheshti, Elisa Bertino, and Norman Foo.
\newblock Reputation management in crowdsourcing systems.
\newblock In {\em 8th International Conference on Collaborative Computing:
  Networking, Applications and Worksharing, CollaborateCom 2012, Pittsburgh,
  PA, USA, October 14-17, 2012}, pages 664--671. {IEEE}, 2012.

\bibitem{DBLP:conf/apweb/AllahbakhshIBBBF13}
Mohammad Allahbakhsh, Aleksandar Ignjatovic, Boualem Benatallah,
  Seyed{-}Mehdi{-}Reza Beheshti, Elisa Bertino, and Norman Foo.
\newblock Collusion detection in online rating systems.
\newblock In Yoshiharu Ishikawa, Jianzhong Li, Wei Wang, Rui Zhang, and Wenjie
  Zhang, editors, {\em Web Technologies and Applications - 15th Asia-Pacific
  Web Conference, APWeb 2013, Sydney, Australia, April 4-6, 2013. Proceedings},
  volume 7808 of {\em Lecture Notes in Computer Science}, pages 196--207.
  Springer, 2013.

\bibitem{DBLP:journals/compsec/AllahbakhshIBBFB14}
Mohammad Allahbakhsh, Aleksandar Ignjatovic, Boualem Benatallah,
  Seyed{-}Mehdi{-}Reza Beheshti, Norman Foo, and Elisa Bertino.
\newblock Representation and querying of unfair evaluations in social rating
  systems.
\newblock {\em Computers {\&} Security}, 41:68--88, 2014.

\bibitem{beheshti2012organizing}
Seyed Mehdi~Reza Beheshti.
\newblock {\em Organizing, Querying, and Analyzing Ad-hoc Processes' Data}.
\newblock PhD thesis, UNIVERSITY OF NEW SOUTH WALES SYDNEY{\textperiodcentered}
  AUSTRALIA, 2012.

\bibitem{DBLP:conf/caise/BeheshtiBN13}
Seyed{-}Mehdi{-}Reza Beheshti, Boualem Benatallah, and Hamid R.~Motahari
  Nezhad.
\newblock Enabling the analysis of cross-cutting aspects in ad-hoc processes.
\newblock In Camille Salinesi, Moira~C. Norrie, and Oscar Pastor, editors, {\em
  Advanced Information Systems Engineering - 25th International Conference,
  CAiSE 2013, Valencia, Spain, June 17-21, 2013. Proceedings}, volume 7908 of
  {\em Lecture Notes in Computer Science}, pages 51--67. Springer, 2013.

\bibitem{DBLP:conf/wise/BeheshtiBNA12}
Seyed{-}Mehdi{-}Reza Beheshti, Boualem Benatallah, Hamid R.~Motahari Nezhad,
  and Mohammad Allahbakhsh.
\newblock A framework and a language for on-line analytical processing on
  graphs.
\newblock In Xiaoyang~Sean Wang, Isabel~F. Cruz, Alex Delis, and Guangyan
  Huang, editors, {\em Web Information Systems Engineering - {WISE} 2012 - 13th
  International Conference, Paphos, Cyprus, November 28-30, 2012. Proceedings},
  volume 7651 of {\em Lecture Notes in Computer Science}, pages 213--227.
  Springer, 2012.

\bibitem{DBLP:conf/bpm/BeheshtiBNS11}
Seyed{-}Mehdi{-}Reza Beheshti, Boualem Benatallah, Hamid R.~Motahari Nezhad,
  and Sherif Sakr.
\newblock A query language for analyzing business processes execution.
\newblock In Stefanie Rinderle{-}Ma, Farouk Toumani, and Karsten Wolf, editors,
  {\em Business Process Management - 9th International Conference, {BPM} 2011,
  Clermont-Ferrand, France, August 30 - September 2, 2011. Proceedings}, volume
  6896 of {\em Lecture Notes in Computer Science}, pages 281--297. Springer,
  2011.

\bibitem{DBLP:conf/springsim/BeheshtiM07}
Seyed{-}Mehdi{-}Reza Beheshti and Mohsen~Sadighi Moshkenani.
\newblock Development of grid resource discovery service based on semantic
  information.
\newblock In George~F. Riley, editor, {\em Proceedings of the 2007 Spring
  Simulation Multiconference, SpringSim 2007, Norfolk, Virginia, USA, March
  25-29, 2007, Volume 1}, pages 141--148. {SCS/ACM}, 2007.

\bibitem{DBLP:journals/corr/abs-1211-5009}
Seyed{-}Mehdi{-}Reza Beheshti, Hamid R.~Motahari Nezhad, and Boualem
  Benatallah.
\newblock Temporal provenance model {(TPM):} model and query language.
\newblock {\em CoRR}, abs/1211.5009, 2012.

\bibitem{DBLP:journals/corr/abs-1211-5817}
Seyed{-}Mehdi{-}Reza Beheshti, Sherif Sakr, Boualem Benatallah, and Hamid
  R.~Motahari Nezhad.
\newblock Extending {SPARQL} to support entity grouping and path queries.
\newblock {\em CoRR}, abs/1211.5817, 2012.

\bibitem{hull2008artifact}
Richard Hull.
\newblock Artifact-centric business process models: Brief survey of research
  results and challenges.
\newblock In {\em On the Move to Meaningful Internet Systems: OTM 2008}, pages
  1152--1163. Springer, 2008.

\bibitem{hull2011formal}
Richard Hull, Elio Damaggio, Riccardo De~Masellis, Fabiana Fournier, Manmohan
  Gupta, III Fenno Terry~Heath, Stacy Hobson, Mark Linehan, Sridhar Maradugu,
  Anil Nigam, et~al.
\newblock A formal introduction to business artifacts with
  guard-stage-milestone lifecycles.
\newblock 2011.

\bibitem{hull2011business}
Richard Hull, Elio Damaggio, Riccardo De~Masellis, Fabiana Fournier, Manmohan
  Gupta, Fenno~Terry Heath~III, Stacy Hobson, Mark Linehan, Sridhar Maradugu,
  Anil Nigam, et~al.
\newblock Business artifacts with guard-stage-milestone lifecycles: managing
  artifact interactions with conditions and events.
\newblock In {\em Proceedings of the 5th ACM international conference on
  Distributed event-based system}, pages 51--62. ACM, 2011.

\bibitem{hull2011introducing}
Richard Hull, Elio Damaggio, Fabiana Fournier, Manmohan Gupta, Fenno~Terry
  Heath~III, Stacy Hobson, Mark Linehan, Sridhar Maradugu, Anil Nigam,
  Piyawadee Sukaviriya, et~al.
\newblock Introducing the guard-stage-milestone approach for specifying
  business entity lifecycles.
\newblock In {\em Web Services and Formal Methods}, pages 1--24. Springer,
  2011.

\bibitem{hull2009facilitating}
Richard Hull, Nanjangud~C Narendra, and Anil Nigam.
\newblock Facilitating workflow interoperation using artifact-centric hubs.
\newblock In {\em Service-Oriented Computing}, pages 1--18. Springer, 2009.

\bibitem{lohmann2013compliance}
Niels Lohmann.
\newblock Compliance by design for artifact-centric business processes.
\newblock {\em Information Systems}, 38(4):606--618, 2013.

\bibitem{nandi2005adaptive}
Prabir Nandi and Santhosh Kumaran.
\newblock Adaptive business objects-a new component model for business
  integration.
\newblock In {\em ICEIS (3)}, pages 179--188, 2005.

\bibitem{ram2009new}
Sudha Ram and Jun Liu.
\newblock A new perspective on semantics of data provenance.
\newblock {\em SWPM}, 526, 2009.

\bibitem{wang2005framework}
Jianrui Wang and Akhil Kumar.
\newblock A framework for document-driven workflow systems.
\newblock In {\em Business Process Management}, pages 285--301. Springer, 2005.

\end{thebibliography}

\end{document}